\documentclass{nature}

\usepackage[margin=0.7in]{geometry}

\usepackage{graphicx}
\usepackage{amssymb}
\usepackage{mathtools}
\usepackage{xfrac}

\usepackage{footnote}

\usepackage{color}

\definecolor{red}{rgb}{1,0,0}

\title{A unified model of ripples and dunes in water and planetary environments}

\author{Orencio Duran Vinent$^{1,2}$, Bruno Andreotti$^{3}$, Philippe Claudin$^3$ \& Christian Winter$^{1,4}$}

\begin{document}

\maketitle

\begin{affiliations}
\item MARUM Center for Marine Environmental Sciences, Bremen University, Leobener Str., D-28359 Bremen, Germany
\item Department of Ocean Engineering, Texas A\&M University, College Station, TX 77843-3136, USA.
\item Laboratoire de Physique et M\'ecanique des Milieux H\'et\'erog\`enes, PMMH UMR 7636 ESPCI -- CNRS -- Univ. Paris-Diderot -- Univ. P.M. Curie, 10 rue Vauquelin, 75005 Paris, France
\item Institute of Geosciences, Christian-Albrechts University Kiel, Otto-Hahn Pl. 1, D-24116 Kiel, Germany
\end{affiliations}

\begin{abstract}
Subaqueous and aeolian bedforms are ubiquitous on Earth and other planetary environments. However, it is still unclear which hydrodynamical mechanisms lead to the observed variety of morphologies of self-organized natural patterns such as ripples, dunes or compound bedforms. Here we present simulations with a coupled hydrodynamic and sediment transport model that resolve the initial and mature stages of subaqueous and aeolian bedform evolution in the limit of large flow thickness. We identify two types of bedforms consistent with subaqueous ripples and dunes, and separated by a gap in wavelength. This gap is explained in terms of an anomalous hydrodynamic response in the structure of the inner boundary layer that leads to a shift of the position of the maximum shear stress from upstream to downstream of the crest. This anomaly gradually disappears when the bed becomes hydrodynamically rough. By also considering the effect of the spatial relaxation of sediment transport we provide a new unifying framework to compare ripples and dunes in planetary environments to their terrestrial counterparts.
\end{abstract}

Bedforms like aeolian dunes, found in arid regions and along shorelines, and subaqueous ripples and dunes in riverine and marine environments, result from the same instability mechanism, related to the lag between bed shear stress, sediment transport and bed elevation \cite{sauermann_continuum_2001,kroy_minimal_2002,andreotti_selection_2002,fourriere_bedforms_2010,colombini_ripple_2011,duran_aeolian_2011,charru_sand_2013}. Analogous bedforms are also observed in other planetary environments as diverse as Mars, Venus, Titan or comet Churyumov-Gerasimenko \cite{bourke_extraterrestrial_2010,diniega_our_2017,jia_giant_2017}. They can display, however, rather different characteristic lengths and formation time scales. In order to deduce a reliable interpretation of these atmospheric conditions from remote photos and measurements, an obstacle is to determine relevant terrestrial analogues. Martian meter scale bedforms were for instance first understood as aeolian ripples superimposed on barchan dunes \cite{silvestro_ripple_2010,duran_direct_2014}, but have been recently reinterpreted as being more akin to subaqueous current ripples \cite{silvestro_dune-like_2016,lapotre_large_2016}. But, how could aeolian-like bedforms (dunes and impact ripples\cite{duran_direct_2014}) and subaqueous-like bedforms (current ripples) coexist in the same Martian environment? Another source of confusion in the literature results from the analysis of the pattern wavelength -- defined as the average crest to crest distance. Ripples usually refer to bedforms of small dimension, typically scaling \cite{yalin_determination_1985,ashley_classification_1990} with the grain size $d$, whereas dunes are associated with larger dimensions, with a wavelength on the order of the river depth in the subaqueous case \cite{ashley_classification_1990,charru_sand_2013} or that can reach the thickness of the convective atmospheric boundary layer in the aeolian one \cite{andreotti_giant_2009}. However, this classification must be revisited when planetary bedforms are investigated: `small' or `large' can mean different actual sizes when the hydrodynamic parameters take unusual values as compared to terrestrial aeolian or subaqueous environments. How should we name, for instance, the decimeter scale bedforms emerging in the wind tunnel reproducing the high CO$_2$ pressure of Venusian atmospheric conditions \cite{greeley_microdunes_1984}? Furthermore, the role of the slow increase of the bedform wavelength by pattern coarsening is still debated \cite{fourriere_bedforms_2010}. Are subaqueous dunes the result of such an evolution? Are ripples and dunes fundamentally different bedform types? Here we propose a unified model that describes the full dynamics of bedforms from initiation to saturation. This opens the way of solving the inverse problem of deriving hydrodynamic conditions from the study of bedform patterns in planetary environments.

\section*{Simulation of subaqueous bedforms}

The physics-based model describes the interaction between bed elevation, which modulates the flow and therefore the bed shear stress, and sediment transport, which controls erosion and deposition (Methods, see also Supp.~Information and Supp.Figs.~1-8 for model calibration and validation). The hydrodynamic part, based on the linear response of the flow to the topographic relief, accounts for two effects previously ignored in similar dune models\cite{kroy_minimal_2002,andreotti_selection_2002,schwammle_model_2005} but recently emphasized as being relevant for pattern formation by the dissolution instability\cite{claudin_dissolution_2017}. One is the hydrodynamic roughness induced by the granular bed or by sediment transport. The other is the shift of the position of the maximum bed shear stress from upstream to downstream of the crest (an effect we refer to as a `hydrodynamic anomaly' in what follows) associated with a turbulent transition of the topographic response. 

The typical temporal evolution of a sediment bed resulting from numerical simulations of the model in the subaqueous case is illustrated in Fig.~1 (see also Supplementary Movies 1, 2 and 3 and Supp.Fig.~9) for different values of the grain-based Reynolds number $\mathcal{R}_d = du_*/\nu$, where $d$ is the grain size, $u_*$ is the shear velocity of the flow and $\nu$ its kinematic viscosity. Starting from a flat bed with a very low amount of random noise, ripples first emerge at small wavelength, in agreement with the predictions of the linear stability analysis of the problem. Ripples then grow in height and quickly reach the aspect ratio at which the turbulent boundary layer separates from the bed, forming a recirculation bubble and an avalanche slip face. The space-time diagrams show that the migration and merging of interacting ripples leads to pattern coarsening associated with an increase of the wavelength $\lambda$. This evolution is slow enough for the bedform height to quickly adapt to its wavelength, maintaining a quasi-constant aspect ratio (Supp.Fig.~10) -- i.e. bedform height is not a variable independent of $\lambda$. Depending on the parameters ($\mathcal{R}_d$ in Fig.~1), the final pattern corresponds to mature ripples, smooth dunes or dunes with superimposed ripples, as observed in natural and laboratory conditions. These numerical simulations accurately describe the complexity of bedform evolution in subaqueous experiments\cite{venditti_morphodynamics_2005,venditti_interfacial_2006} and quantitatively reproduce the characteristics of subaqueous and aeolian ripples and dunes (e.g. Supp.Fig.~10, Fig.~2 and Supp.Figs.~11-16). Hence, they provide a reliable physical framework to study bedforms also in other planetary environments for which data sets are more sparse or incomplete and require interpretation.

\section*{Hydrodynamic instability and bedform initiation}

The physical processes leading to the emergence of bedforms can be deduced from the linear stability analysis of a flat sand bed, i.e. in the limit of vanishing bedform height. The instability mechanism at work directly follows from hydrodynamics (Fig.~3a). For an initial perturbation of a flat bed, like a sediment hump, flow streamlines converge along the upstream side of the crest, increasing flow velocity, while they diverge on the downstream side. As the sediment flux increases with the flow velocity, this results in erosion of sediments at the upstream side and subsequent deposition at the downstream side, i.e. a propagation of the hump. For bedforms to grow, the position of the maximum bed shear stress must be upstream of the crest by a distance $\delta_\tau > 0$, which is a necessary condition for sediment deposition around the crest (see Methods and Fig.~3a), as already recognized in pioneering works\cite{richards_formation_1980,MCLEAN1990131}. This spatial shift $\delta_\tau$ is thus the key quantity that characterizes hydrodynamics around bedforms.

Former modeling approaches consider a relatively constant upstream position of the maximum bed shear stress relative to the bedform wavelength\cite{MCLEAN1990131,kroy_minimal_2002,andreotti_selection_2002,schwammle_model_2005}. However, our hydrodynamic model\cite{claudin_dissolution_2017} shows that the spatial shift $\delta_\tau$ depends on both, a Reynolds number $\mathcal{R}_\lambda = \lambda u_*/\nu$ based on the bed wavelength $\lambda$, and the grain-based Reynolds number $\mathcal{R}_d$. Indeed, the position of the maximum bed shear stress is inextricably linked to the state of the inner boundary layer (from viscous to inertial to turbulent dominated regimes as $\mathcal{R}_\lambda = \lambda u_*/\nu$ increases) which differs depending on the hydrodynamic roughness of the bedform surface, as dictated by $\mathcal{R}_d$ (Fig.~2).

At large and small wavelengths, where the hydrodynamic response is dominated by either inertia (advection) or turbulent diffusion (Reynolds stress) the maximum shear stress is located upstream of the crest ($\delta_\tau>0$), so that bedforms of size $\lambda$ can grow. However, for transitional wavelengths, the maximum bed shear stress shifts downstream of the crest ($\delta_\tau < 0$) and bedforms eventually erode and decay. This hydrodynamic anomaly is associated with a strong modulation of the turbulent intensity as the disturbance induced by the topography remains laminar on the upstream side of a hump but gradually becomes turbulent on the downstream side due to the amplification of turbulent fluctuations when streamlines diverge \cite{charru_sand_2013,claudin_dissolution_2017} (Fig.~2a). The hydrodynamic anomaly gradually disappears at the transition from an hydrodynamically smooth to an hydrodynamically rough bed ($\mathcal{R}_d \gtrsim  20$) as the thickness of the viscous sub-layer becomes smaller than the grain or the transport layer size and the rough surface generates turbulent fluctuations everywhere (Fig.~2). The maximum bed shear stress thus shifts from downstream to upstream of the crest and bedform growth is recovered.

\section*{Formation of subaqueous ripples and dunes}

Simulations of subaqueous conditions show that the formation of ripples from linear instability in the smooth regime (Fig~2, solid lines) is followed by pattern coarsening that eventually saturates (Fig~2, `saturated bedforms' line) once their wavelength approaches the hydrodynamic anomaly at the end of the transitional range of $\mathcal{R}_\lambda$. Dunes can then emerge at long times, which, contrarily to ripples, have a turbulent dominated hydrodynamic response (Fig.~2, `superimposed bedforms' line). These dunes continuously grow by pattern coarsening while keeping a complex array of superimposed ripples, whose hydrodynamical response is either transitional or dominated by inertia. These findings match observations as e.g. made on the Leyre river \cite{fourriere_bedforms_2010}, which show the pattern coarsening, the stabilisation of ripple size and the later transition to dunes with superimposed ripples (Supp.Fig.~9). Finally, in the rough regime the hydrodynamic anomaly disappears and, as $\mathcal{R}_d$ increases, the initial wavelength presents a discontinuous transition from ripples to dunes. These smooth dunes continuously growth by pattern coarsening.

The simulations quantitatively reproduce the range of measurements reported from controlled flume experiments (see Supp.Fig.~13 for data sources and Supp.Figs.~11-15 for additional comparisons) and allow us to identify subaqueous ripples with an inertial or transitional hydrodynamic response and subaqueous dunes with a turbulent hydrodynamic response (Fig.~2). The data is also consistent with the three final bedform types investigated in this article: growing dunes for rough flows (Fig.~1c), growing dunes with superimposed (stable) ripples for smooth and intermediate flows (Fig.~1b) and stable mature ripples at small $\mathcal{R}_d$ (Fig.~1a).

\section*{Effect of transport saturation length}

A complete understanding of the bedform phase diagram also needs to consider the influence of the spatial relaxation of sediment transport on bedform growth. As grains adjust to the flow drag, the sediment transport rate spatially relaxes towards equilibrium with the local shear stress. This transient is characterised by the saturation length $l_{\rm sat}$ \cite{sauermann_continuum_2001,andreotti_selection_2002,andreotti_measurements_2010}, which defines the space lag between shear stress and sediment flux. Consequently, the position of the maximum sediment flux relative to a hump crest is shifted a distance $l_{\rm sat}$ downstream the maximum bed shear stress (Methods). Because sediment deposition occurs downstream of the maximum sediment flux, a hump only grows if the maximum sediment flux is located upstream of the crest, which leads to the condition $\delta_\tau > l_{\rm sat}$ (Fig.~3a). This condition naturally imposes a minimum wavelength for bedforms as the shear stress spatial shift $\delta_\tau$ scales with $\lambda$ (see Methods for a derivation based on linear stability analysis). 

Transport relaxation thus acts as a stabilising mechanism for short wavelengths and essentially controls the initial bedform wavelength. For smooth and transitional flows, when the initial wavelength imposed by the saturation length is larger than the size at which mature ripples saturates, only dunes are possible. Otherwise, both ripples and dunes can form and coexist (Fig.~3b and Fig.~4a, see Supp.Fig.~16 for data sources). For hydrodynamically rough flows, there is no distinction between ripples and dunes and the initial wavelength increases continuously with $l_{\rm sat}$ (Fig~3c). Similarly, for a given saturation length, the initial wavelength presents a discontinuity in the transition from smooth to hydrodynamically rough flows as the ripple mode disappears (the stable region $\delta_\tau < l_{\rm sat}$ can be inferred from Fig.~2).

Our theory can also be tested against data in the case of aeolian sediment transport in saltation, for which $l_{\rm sat}$  has been directly and independently measured and modeled (Supp.Fig.~6). Contrarily to the subaqueous case, the hydrodynamic roughness increases with sediment transport, so that $\mathcal{R}_d$ must be corrected to include this effect (see Methods). Aeolian dunes are found to be in the rough hydrodynamic regime and are analogous to subaqueous dunes without superimposed ripples. The wavelength at which these aeolian dunes emerge (Fig.~4b) is in quantitative agreement with the results of the simulations, thus giving strong support for the model.

\section*{Interpretation of planetary bedforms}  

Having quantitatively validated the model on subaqueous and aeolian bedforms, we can use it to analyse less understood situations such as  planetary bedforms on Venus and Mars (Fig.~4c,d). Ignoring for simplicity effects due to confinement by a free surface (as in rivers) or by a capping inversion layer (in the atmosphere), the phase diagram of bedforms for monodisperse sand in unidirectional flows depends on two dimensionless parameters: the grain-based Reynolds number $\mathcal{R}_d$, which controls the turbulent roughening transition and the saturation length rescaled by the size of the viscous sublayer $l_{\rm sat} u_*/\nu $, which determines the initial wavelength. The resulting diagram, displayed in Fig.~5, shows that data for subaqueous bedforms dunes match the theoretical predictions (coloured area). The emergence of ripples, dunes, or dunes with superimposed ripples is controlled, in a first approximation, by two different thresholds (Fig.~5): the smooth--rough transition in the disturbed flow at $\mathcal{R}_d \approx 20$, and the stabilizing effect of transport relaxation on ripples for $l_{\rm sat} u_*/\nu \approx 10^3$. Interestingly, the existing relation between $l_{\rm sat}/d$, $\mathcal{R}_d$ and the Galileo number (see Methods) allows one to recast the bedform diagram (Fig.~5) in parameter spaces unrelated to the saturation length. In particular, it explains the most common phenomenological diagram represented in terms of the Shields number and dimensionless grain diameter\cite{berg_new_1993} (Supp.Fig.~17). 

Considering only the wavelength based Reynolds numbers $\mathcal{R}_\lambda$ in Fig.~4, we find that mid-size Martian bedforms are analogous to subaqueous ripples, as recently suggested by the analysis of their morphology~\cite{lapotre_large_2016}. Mid-size Martian bedforms, as defined in Fig.~4d, include large ripples, largely found in monodisperse sand but also reported in polydisperse sediments~\cite{lapotre_large_2016,lapotre_morphologic_2018,weitz_sand_2018}, and TARs (transversal aeolian ridges) that seems to form mainly on polydisperse sediments~\cite{hugenholtz_morphology_2017,geissler_morphology_2017}. This highlights the robustness of the hydrodynamic mechanism stabilising smaller-than-dune bedforms, which is independent on the details of sediment transport and grain size distribution. In contrast, although some small Venusian ripples (obtained in a high pressure wind tunnel\cite{greeley_microdunes_1984}) could be analogous to subaqueous ripples, most Venusian bedforms are consistent with growing dunes without stable ripples, as reported in the experiments\cite{greeley_microdunes_1984}. Assuming that the smallest wavelength of these bedforms corresponds to the most unstable mode, one can determine the corresponding saturation length $l_{\rm sat}$ and thus complete the phase diagram in Figs.~4 and 5 (see Methods for further details on the calculation of $l_{\rm sat}$ for Mars). Bedforms at rather low $\mathcal{R}_\lambda$ emerge in both Martian and water environments, in spite of their very different conditions, because the very low density atmosphere and large viscous sublayer in Mars, as compared to water, compensate for the large saturation length for saltation on Mars, of the order of $10^3-10^4d$ (Methods) that is much larger than in water ($l_{\rm sat} \sim 10d-100d$, see Supp.Fig.~5). 

Martian conditions are thus analogous to the subaqueous regime where ripples superimpose to dunes (Fig.~4). Like ripples in water, the scale of mid-size bedforms on Mars is selected by the thickness of the viscous sublayer $\nu/u_*$. Interestingly, our model predicts that Martian dunes also scale with $\nu/u_*$ (Fig.~4d) but with a much larger prefactor, in contrast to dunes on Earth that scale with the transport saturation length $l_{\rm sat}$\cite{andreotti_selection_2002,claudin_scaling_2006,sauermann_continuum_2001} (Fig.~4b and Supp.Fig.~18). The phase diagram can be further completed by adding a semi-empirical estimation of the scaling of impact ripples in monodisperse sand (Methods). Impact ripples are the smallest bedforms and do not have a hydrodynamic origin but result from the interaction of saltation transport with the bed elevation\cite{duran_direct_2014}. On Earth and Mars (see Methods), impact ripples show a clear scale separation with the other bedforms (Fig.~4), which means they are ubiquitous and can appear superimposed to any other bedform type\cite{lapotre_large_2016,sullivan_wind-driven_2008} (Supp.Fig.~19). Note that in polydisperse sand, another type of larger impact ripples, the so-called megaripples, can form, which has been proposed to be the terrestrial analogue to TARs\cite{bridges_formation_2015,hugenholtz_morphology_2017,geissler_morphology_2017}.

In summary, we identify two types of bedforms created by a hydrodynamic instability in the limit of large flow thickness: those inducing an inertial hydrodynamic response and those inducing a turbulent hydrodynamic response. They are separated by a gap in wavelength explained in terms of an hydrodynamic anomaly that leads to a shift of the position of the maximum shear stress from upstream to downstream of the crest. This anomaly gradually disappears when the bed becomes hydrodynamically rough. These bedform types are consistent with subaqueous ripples and dunes, which then represent a suitable reference to classify bedforms in other environments (Fig.~5, Supp.Figs.~19 and 20, and Supp.Table.~8). The result is robust, regardless the details of sediment transport and bed segregation, as it arises from an hydrodynamic mechanism which has been directly evidenced and characterized in well controlled experiments \cite{abrams_relaxation_1985,frederick_velocity_1988}. For monodisperse sand and non-suspended transport in the limit of large flow thickness, the parameter space physically relevant to determine bedform morphology is based on two strongly correlated dimensionless numbers, $\mathcal{R}_d$ and $l_{\rm sat} u_*/\nu$. The grain-based Reynolds number $\mathcal{R}_d$ encodes the hydrodynamic roughness and controls the appearance of the hydrodynamic anomaly. The rescaled saturation length $l_{\rm sat} u_*/\nu$ depends on the dominant mode of sediment transport and determines the initial wavelength. The occurrence of ripple-like and dune-like bedforms is controlled by two different thresholds: the rough transition at $\mathcal{R}_d \approx 20$, and the stabilising effect of transport relaxation at $l_{\rm sat} u_*/\nu \approx 10^3$. Whenever ripples are present, the size of the viscous sublayer $\nu/u_*$ determines the scale of the maximum size of steady state ripples and the minimum size of emerging dunes. When only dunes are possible, their minimum size scale with the transport saturation length. Our findings provide a unifying framework to understand planetary bedforms and suggests a correspondence between bedforms in different environments based on the formation mechanism.

These results point to several directions for future research. An extension of this work to account for the stabilising role of finite flow depth\cite{fourriere_bedforms_2010} would provide the general bedform diagram, and the possibility to discuss for example the yet unexplained upper plane regime\cite{southard_bed_1990,berg_new_1993,baas_predicting_2016}. Direct measurements of the saturation length in the case of bed load, for different grain size, flow velocity and grain to fluid density ratio, are clearly needed. In particular, the transition from bed load to saltation as $\rho_p/\rho_f$ increases remains unexplored experimentally. The transition to suspended load would also be interesting to consider, as it dramatically increases the saturation length\cite{andreotti_bedforms_2012}. Finally, wind tunnel experiments in low pressure conditions would shed more light on Martian sediment transport, key to understand better the role of grain segregation and size polydispersity in the formation of Martian megaripples and TARs, which need further investigation\cite{hugenholtz_morphology_2017,geissler_morphology_2017,lammel_aeolian_2018}.

\begin{methods}

%________________
\subsection{Morphodynamic model.}

We consider a unidirectional flow of infinite depth, density $\rho_f$, kinematic viscosity $\nu$ and average shear velocity $u_*$ (defined using the average bed shear stress $\rho_f u_*^2$) acting along the $x$-direction on a granular bed of elevation $Z(x,t)$, particle size $d$ and density $\rho_p$. The changes in the surface $Z(x,t)$ are governed by the spatial variations of the sediment flux $q(x)$, which result from the modulation of the bed shear stress $\tau_b(x)$ in response to the flow on a non-flat surface. The mass conservation equation reads
\begin{equation}
\partial_t Z = - \partial_x q,
\end{equation}
where $q$ is the sediment flux defined as volume of sediment, packed at the bed volume fraction $\phi_b$, which is transported per unit transverse length and unit time. The initial condition is a flat bed with very small random noise whose amplitude is comparable to the particle size $d$. We use periodic boundary conditions in the flow direction.

%________________
\subsection{Sediment transport.}

Experiments\cite{andreotti_measurements_2010} show that the sediment flux needs some space to adapt to its equilibrium (or `saturated') value $q_{\rm sat}$. Close to saturation this spatial relaxation is well described by:
\begin{equation}
	\partial_x q = (q_{\rm sat} - q) / l_{\rm sat} 
	\label{qx}
\end{equation}
where $l_{\rm sat}$ is the saturation length\cite{sauermann_continuum_2001,kroy_minimal_2002,andreotti_selection_2002}. As detailed below, we take the saturated flux $q_{\rm sat}$ as a function of the local bed shear stress $\tau_b$. A consequence of Eq.~\ref{qx} is that the sediment flux on a wavy solid bed of wavenumber $k$ is shifted downstream of the saturated flux (or equivalently the bed shear stress) by a spatial lag $k^{-1}\tan^{-1}{(k l_{\rm sat})} \approx l_{\rm sat}$, as sketched in Fig.~3a.

%________________
\subsection{Saturated sediment flux for bedload.}

For bedload transport (typical of dense fluids like water on Earth), we express $q_{\rm sat}(\tau_b)$ using a Meyer-Peter \& M\"uller type of transport law obtained from direct transport simulations\cite{duran_numerical_2012}:
\begin{equation}
	q_{\rm sat} = a_q Q \left( \frac{\tau_b}{\tau_t} - 1 \right) \sqrt{\frac{\tau_b}{\tau_t}},
\end{equation}
where $\tau_t$ is the bed shear stress at the transport cessation threshold. We define the corresponding threshold shear velocity as $u_t = \sqrt{\tau_t / \rho_f}$. The flux scale is $Q = u_t^2 d/\sqrt{(s - 1) g d}$, where $s = \rho_p/\rho_f$ is the particle-fluid-density ratio, $g$ is the gravity acceleration and $a_q$ is a proportionality constant, which, for simplicity, is set to $a_q = 1$, as suggested by transport simulations\cite{duran_numerical_2012}.

%________________
\subsection{Saturated sediment flux for saltation (air).}

For saltation transport (typical of low density fluids like air on Earth) we use an Ungar \& Haff\cite{ungar_steady_1987} type of transport law also obtained from direct transport simulations\cite{duran_numerical_2012}:
\begin{equation}
	q_{\rm sat} = a_q Q \left( \frac{\tau_b}{\tau_t} - 1 \right),
\end{equation}
where $Q = u_t^3/[(s - 1) g]$ and the proportionality constant is $a_q\simeq 1$ according to transport simulations\cite{duran_numerical_2012}.

%________________
\subsection{Saturation length for bedload.}

We indirectly obtained the saturation length $l_{\rm sat}$ by fitting the model to obtain measured wavelengths of initial bedforms (Supp.~Information and Supp.Fig.~4). From dimensional analysis we find $l_{\rm sat}/d$ scales with the grain-based Reynolds number $\mathcal{R}_d = d u_*/\nu$ and the Galileo number $G = d\sqrt{(s-1)gd}/\nu$, and can be fitted by the expression:
\begin{equation}
	l_{\rm sat}/d = a_b \, \mathcal{R}_d / G^{1.2},
	\label{lsat_b}
\end{equation}
with proportionality constant $a_b = 140$ (see Supp.~Information for further details). In Venus wind tunnel experiments\cite{marshall_experimental_1992}, no impact ripples were observed, suggesting that sediment transport ressembles bedload. As for water in Supp.Fig.~4, we measure indirectly $l_{\rm sat}$ from the comparison of the smallest measured bedforms\cite{greeley_microdunes_1984} to model predictions, which gives $l_{\rm sat}/d \approx 10$ (Fig.~4c).

%________________
\subsection{Saturation length for saltation.}

Following previous analytical models of sediment transport\cite{sauermann_continuum_2001,pahtz_flux_2013}, we assume that the saturation length $l_{\rm sat}$ for saltation scales with the average hop length $v^2_x/g$ of particle trajectories, where $v_x$ is the average particle velocity. Numerical simulations\cite{pahtz_fluid_2017} show that $v_x$ is proportional to the impact threshold shear velocity $u_t$, so that
\begin{equation}
	l_{\rm sat}/d = a_s \, u_t^2 / (g d) \approx a_s s \Theta_t ,
	\label{lsat_s}
\end{equation}
where $\Theta_t = u^2_t/[(s-1) gd]$ is the threshold Shields number. We obtain $a_s = 210 \pm 40$ after fitting the model to wind tunnel and field measurements of the saturation length of sand on air\cite{andreotti_measurements_2010} ($s=2200$ and $\Theta_t \simeq 0.01$) as shown in Supp.Fig.~6. When comparing the model to Mars data, the saturation length is calculated from Eq.~\ref{lsat_s} using the impact threshold velocity $u_t$ predicted for typical particles sizes and gravity constant $g$ (Supp. Tables~1,6 and 7). To account for the uncertainty behind Eq.~\ref{lsat_s} when applied to Mars, we recalibrate the parameter $a_s$ for Martian conditions comparing model prediction to recent in-situ measurements\cite{lapotre_large_2016,sullivan_aeolian_2017} (Supp. Tables~5 and 7), which gives $a_s \approx 100$, only a factor two from the value obtained for Earth.

%________________
\subsection{Sediment transport by avalanches.}

For slopes steeper than the angle of repose $\theta = 34^\circ$, we model the quasi-instantaneous surface relaxation due to avalanches by adding a downslope component $q_{\rm aval}$ to the sediment flux: 
\begin{equation}
	q_{\rm aval} = - Q_{\rm aval} \left[ \tanh(|\partial_x Z|) - \tanh(\tan\theta) \right] \partial_x Z/|\partial_x Z|,
\end{equation}
where $Q_{\rm aval}$, which sets the avalanche time scale, is chosen as large as permitted by numerical time and length scales.

%________________
\subsection{Bed shear stress over a non-flat surface.}

We use the formulation in \cite{claudin_dissolution_2017} to calculate the bed shear stress $\tau_b$ exerted by an uni-directional flow over a relief with elevation $z=Z(x)$. The bed shear stress can be written as $\tau_b = \rho_f u_*^2 (1 + \tau)$,
where the Fourier transform $\hat \tau$ of the dimensionless shear stress perturbation $\tau$ is defined as $\hat\tau = k \hat Z (\mathcal{A} + i \mathcal{B})$. $\hat Z(k)$ is the Fourier transform of $Z(x)$, $k = 2 \pi/\lambda$ the wave number and $\lambda$ the wavelength. Both, in-phase and in-quadrature components of the shear stress modulation, $\mathcal{A}$ and $\mathcal{B}$ respectively, are function of the bedform-based Reynolds number $\mathcal{R}_\lambda = \lambda u_*/\nu$ and the grain-based Reynolds number $\mathcal{R}_d = d u_*/\nu$.
For a bedform of wavelength $\lambda$, the spatial shift between the bed shear stress perturbation and the bed is given by 
\begin{equation}
 	\delta_\tau = \frac{\lambda}{2\pi} \tan^{-1}{(\mathcal{B}/\mathcal{A})} .
\end{equation}
Thus, $\delta_\tau / d$ is only function of $\mathcal{R}_\lambda$ and $\mathcal{R}_d$ as shown in Fig.~2. 

%________________
\subsection{Effects of sediment transport on the grain-based Reynolds number.} 

$\mathcal{R}_d$ is a dimensionless number representing the roughness-based Reynolds number and thus it has to be modified to incorporate the increase in hydrodynamic roughness due to sediment transport. As shown in Supp.Fig.~7 (and discussed in the Supp.~Information) bedload transport does not changes the hydrodynamic roughness and thus the definition $\mathcal{R}_d = u_* d/\nu$ is directly applicable. Similarly, for Venus conditions, the negative feedback of sediment transport on the flow is negligible, according to grain-based simulations\cite{duran_numerical_2012} for particle-fluid density ratios $s = 10-100$ and close to the transport threshold. This is also consistent with the lack of impact ripples observed in wind tunnel experiments\cite{marshall_experimental_1992}, which suggests a bedload-like transport. For saltation on Earth, however, the hydrodynamic roughness $z_0$ increases with the ratio $u_*/u_t$ and can be fitted by the expression $z_0 = r d \exp[-7 (u_t/u_* - 1) ]$, where $r d$, with $r=1/30$, is the roughness in the absence of transport (for hydrodynamically rough flows, see Supp.~Information). The apparent grain size is thus $d_{\rm mod} = r^{-1} z_0(u_*/u_t) = d \exp[-7 (u_t/u_* - 1)]$ and the grain-based Reynolds number becomes $\mathcal{R}_d = u_* d \nu^{-1} \exp[-7 (u_t/u_* - 1)]$. For saltation on Mars the situation is more complex obviously because of the lack of data, but also because the low $\mathcal{R}_d$ at the transport onset ($\sim 0.1$, Supp. Table~7) suggests saltation takes place in a hydrodynamically smooth flow, a poorly understood regime very different than on Earth. We thus estimate the effect of transport on $\mathcal{R}_d$ assuming most transport on Mars occurs relatively close to the transport threshold ($u_* < 2 u_t$)\cite{sullivan_aeolian_2017}, and using the correction obtained for saltation on Earth. This gives $d_{\rm mod} \lesssim 30 d$ and grain-based Reynolds number is in the range $(u_* d / \nu < \mathcal{R}_d \lesssim 30 u_* d / \nu)$. 

%________________
\subsection{Correction to account for flow separation.}

The model used to calculate $\tau_b$ is only rigorous in the limit of small aspect ratios ($\partial_x Z \lesssim 0.3$)) and therefore without any flow separation and recirculation bubble. For a downslope steeper than $\sim 20^\circ$ the turbulent boundary layer separates. Following standard morphodynamic models\cite{kroy_minimal_2002,andreotti_selection_2002}, we then use a phenomenological approximation that extends the bed surface $Z(x)$ to also encompass the flow recirculation region, from the flow separation point $x_s$ to the reattachment point downstream. We consider $\tau_b = 0$ within this surface extension. For simplicity, we use a parabolic extension $Z_e(x)$ defined by three conditions: a continuous and smooth surface at the separation point, $Z(x_s) = Z_e(x_s)$ and $\partial_x Z|_{x_s} = \partial_x Z_e|_{x_s}$, and the condition $Z_e(x_b + 6 H_s) = Z(x_b)$, where $H_s = Z(x_s) - Z(x_b)$ is the height of the separation region and $x_b$ is the base of the separation region, i.e. the  first point with downslope flatter than $20^\circ$ ($x_b > x_s$). The latter condition ensures a constant reattachment length equal to $6 H_s$ for a flat reattachment area. We also constrain the maximum height of the parabolic extension to about 1.3 $H_s$ to prevent unrealistically large distortions of the bed shear stress close to the brink. We have checked that the results of the model are robust with respect to these numbers. 
Crucially, our linear hydrodynamic model is still applicable to mature bedforms as their aspect ratios are between $0.05$ and $0.2$ (Supp.Fig.~10).

%________________
\subsection{Model validation.}

The validation of the hydrodynamic and transport model is discussed in detail in the Supp.~Information.

%________________
\subsection{Calculation of bedform wavelength and amplitude.}

For given values of $\mathcal{R}_d$ and $l_{\rm sat}/d$, we compute the bed evolution $Z(x,t)$. We calculate the wavelength $\lambda(t)$ as the location of the first local maximum of the autocorrelation function $C(\ell,t) = \int_0^{\infty} Z(x,t) Z(x+\ell,t) {\rm d}x$, with corresponding amplitude $H(t) = 2\sqrt{2C(0,t)}$. 
The function $\lambda(t,\mathcal{R}_d,l_{\rm sat}/d)$ is used to characterize the model predictions in Figs.~2, 4 and Supp.Fig.~9.

%________________
\subsection{Linear instability analysis.}

For bed shear stresses well above the threshold ($\tau_b \gtrsim 2 \tau_t$), the linear instability analysis predicts the growth rate $\sigma$ of any surface perturbation as a function of its wavelength $\lambda = 2 \pi / k$ \cite{fourriere_bedforms_2010}: $\sigma \approx \frac{k^2 Q}{1 - (k l_{\rm sat})^2} (\mathcal{B} - \mathcal{A} k l_{\rm sat})$. Thus, a bedform grows if $\mathcal{B}/\mathcal{A} > k l_{\rm sat} > 0$. This condition can be written in terms of the shear stress shift as: $\delta_{\tau} = k^{-1} \tan^{-1}(\mathcal{B}/\mathcal{A}) > k^{-1} \tan^{-1}(k l_{\rm sat})$, which can be approximated as $\delta_{\tau} \gtrsim l_{\rm sat}$, as discussed in Fig.~3.

%________________
\subsection{Model limitations when comparing with data.}

Although the extension of the model to finite flow depth and finite Froude numbers is straightforward, we focus here on bedforms much smaller than the flow depth, assuming besides an unlimited sand supply. The formation of antidunes, bars or finger dunes is therefore left aside\cite{andreotti_bedforms_2012,baas_predicting_2016,pont_two_2014}. In a further effort of reductionism, we assume unidirectionality of the flow, monodisperse sediments, and the absence of vegetation. Accordingly, the results of the numerical simulations will be compared to field or experimental measurements where these conditions are satisfied.

\subsection{Comparison with data.}

The physical characteristics of given bedforms can be characterised by four length scales, the wavelength $\lambda$, the viscous length $\nu/u_*$, the grain size $d$ and the saturation length $l_{\rm sat}$, which give three dimensionless numbers: the bedform-based Reynolds number $\mathcal{R}_\lambda = \lambda u_*/\nu$, the grain-based Reynolds number $\mathcal{R}_d = d u_*/\nu$ and the rescaled saturation length $l_{\rm sat} u_*/\nu$. The sources of subaqueous experimental and field data shown in Figs.~2 (e.g. \cite{yalin_determination_1985,guy_summary_1966,fourriere_bedforms_2010,martin_origin_2013,baas_flume_1994,baas_empirical_1999,langlois_initiation_2007,coleman_initiation_1996,kuru_formation_1995,jain_growth_1971,jain_spectral_1974,nakagawa_hiroji_spectral_1984}) and 4 (e.g. \cite{yalin_determination_1985,guy_summary_1966,bridge_flow_1988,baas_flume_1994,baas_empirical_1999,venditti_morphodynamics_2005,langlois_initiation_2007,fourriere_bedforms_2010,martin_origin_2013,kuru_formation_1995,coleman_initiation_1996,jain_growth_1971,jain_spectral_1974,nakagawa_hiroji_spectral_1984}) are detailed in Supplementary Figures ~11-16. Data for air on Earth, Venus and Mars, shown in Fig.~4, as well as the parameters used for the calculation of the different quantities, are summarised in Supplementary Tables~1-7 (e.g. \cite{andreotti_measurements_2010,greeley_microdunes_1984,marshall_experimental_1992,greeley_windblown_1984,lapotre_large_2016,sullivan_wind-driven_2008,kok_difference_2010,sullivan_aeolian_2017,newman_winds_2017}). Main sources of errors are discussed in the Supplementary Information.

%________________
\subsection{Wavelength of impact ripples.}
Impact ripples are the smallest bedforms and do not have an hydrodynamic origin but result from the interaction of saltation transport with the bed elevation. Impact ripples are characteristic of saltation and do not appear for bedload sediment transport. The initial wavelength of impact ripples formed in low-density fluids and monodisperse sand scales as\cite{duran_direct_2014} $\lambda \propto s^{-1/2} g^{-1} u_t^2 (u_*/u_t - 1)$, where $s$ is the grain to fluid density ratio. Assuming that the final wavelength is proportional to the initial wavelength, as suggested by wind tunnel measurements\cite{andreotti_aeolian_2006}, we get a semi-empirical estimation of the final wavelength as $\lambda = b s^{-1/2} g^{-1} u_t^2 (u_*/u_t - 1)$, where the constant $b \approx 3200$ is obtained from the best fit of data obtained for aeolian ripples \cite{andreotti_aeolian_2006}. The predicted size of impact ripples for monodisperse sand are added to Fig.~4 using the parametrization of the saturation length $l_{\rm sat}$ for saltation in terms of $u_t$ (Eq.~\ref{lsat_s}). For Earth, we use a constant ratio $u_*/u_t \approx 2.5$, which seems to define the largest reported ripples for monodisperse sand\cite{andreotti_aeolian_2006}. For Mars, we use a constant value $u_* = 0.5$m/s characterizing typical wind conditions\cite{sullivan_aeolian_2017}.

\end{methods}

%\bibliography{bedforms}

\begin{thebibliography}{10}
\expandafter\ifx\csname url\endcsname\relax
  \def\url#1{\texttt{#1}}\fi
\expandafter\ifx\csname urlprefix\endcsname\relax\def\urlprefix{URL }\fi
\providecommand{\bibinfo}[2]{#2}
\providecommand{\eprint}[2][]{\url{#2}}

\bibitem{sauermann_continuum_2001}
\bibinfo{author}{Sauermann, G.}, \bibinfo{author}{Kroy, K.} \&
  \bibinfo{author}{Herrmann, H.~J.}
\newblock \bibinfo{title}{Continuum saltation model for sand dunes}.
\newblock \emph{\bibinfo{journal}{Physical Review E}}
  \textbf{\bibinfo{volume}{64}}, \bibinfo{pages}{031305}
  (\bibinfo{year}{2001}).

\bibitem{kroy_minimal_2002}
\bibinfo{author}{Kroy, K.}, \bibinfo{author}{Sauermann, G.} \&
  \bibinfo{author}{Herrmann, H.~J.}
\newblock \bibinfo{title}{Minimal {Model} for {Sand} {Dunes}}.
\newblock \emph{\bibinfo{journal}{Physical Review Letters}}
  \textbf{\bibinfo{volume}{88}}, \bibinfo{pages}{054301}
  (\bibinfo{year}{2002}).

\bibitem{andreotti_selection_2002}
\bibinfo{author}{Andreotti, B.}, \bibinfo{author}{Claudin, P.} \&
  \bibinfo{author}{Douady, S.}
\newblock \bibinfo{title}{Selection of dune shapes and velocities {Part} 2: {A}
  two-dimensional modelling}.
\newblock \emph{\bibinfo{journal}{The European Physical Journal B - Condensed
  Matter and Complex Systems}} \textbf{\bibinfo{volume}{28}},
  \bibinfo{pages}{341--352} (\bibinfo{year}{2002}).

\bibitem{fourriere_bedforms_2010}
\bibinfo{author}{Fourri\`ere, A.}, \bibinfo{author}{Claudin, P.} \&
  \bibinfo{author}{Andreotti, B.}
\newblock \bibinfo{title}{Bedforms in a turbulent stream: formation of ripples
  by primary linear instability and of dunes by nonlinear pattern coarsening}.
\newblock \emph{\bibinfo{journal}{Journal of Fluid Mechanics}}
  \textbf{\bibinfo{volume}{649}}, \bibinfo{pages}{287--287}
  (\bibinfo{year}{2010}).

\bibitem{colombini_ripple_2011}
\bibinfo{author}{Colombini, M.} \& \bibinfo{author}{Stocchino, a.}
\newblock \bibinfo{title}{Ripple and dune formation in rivers}.
\newblock \emph{\bibinfo{journal}{Journal of Fluid Mechanics}}
  \textbf{\bibinfo{volume}{673}}, \bibinfo{pages}{121--131}
  (\bibinfo{year}{2011}).

\bibitem{duran_aeolian_2011}
\bibinfo{author}{Dur\'an, O.}, \bibinfo{author}{Claudin, P.} \&
  \bibinfo{author}{Andreotti, B.}
\newblock \bibinfo{title}{On aeolian transport: {Grain}-scale interactions,
  dynamical mechanisms and scaling laws}.
\newblock \emph{\bibinfo{journal}{Aeolian Research}}
  \textbf{\bibinfo{volume}{3}}, \bibinfo{pages}{243--270}
  (\bibinfo{year}{2011}).

\bibitem{charru_sand_2013}
\bibinfo{author}{Charru, F.}, \bibinfo{author}{Andreotti, B.} \&
  \bibinfo{author}{Claudin, P.}
\newblock \bibinfo{title}{Sand {Ripples} and {Dunes}}.
\newblock \emph{\bibinfo{journal}{Annual Review of Fluid Mechanics}}
  \textbf{\bibinfo{volume}{45}}, \bibinfo{pages}{469--493}
  (\bibinfo{year}{2013}).

\bibitem{bourke_extraterrestrial_2010}
\bibinfo{author}{Bourke, M.~C.} \emph{et~al.}
\newblock \bibinfo{title}{Extraterrestrial dunes: {An} introduction to the
  special issue on planetary dune systems}.
\newblock \emph{\bibinfo{journal}{Geomorphology}}
  \textbf{\bibinfo{volume}{121}}, \bibinfo{pages}{1--14}
  (\bibinfo{year}{2010}).

\bibitem{diniega_our_2017}
\bibinfo{author}{Diniega, S.} \emph{et~al.}
\newblock \bibinfo{title}{Our evolving understanding of aeolian bedforms, based
  on observation of dunes on different worlds}.
\newblock \emph{\bibinfo{journal}{Aeolian Research}}
  \textbf{\bibinfo{volume}{26}}, \bibinfo{pages}{5--27} (\bibinfo{year}{2017}).

\bibitem{jia_giant_2017}
\bibinfo{author}{Jia, P.}, \bibinfo{author}{Andreotti, B.} \&
  \bibinfo{author}{Claudin, P.}
\newblock \bibinfo{title}{Giant ripples on comet 67p/{Churyumov}-{Gerasimenko}
  sculpted by sunset thermal wind}.
\newblock \emph{\bibinfo{journal}{Proceedings of the National Academy of
  Sciences}} \textbf{\bibinfo{volume}{114}}, \bibinfo{pages}{2509--2514}
  (\bibinfo{year}{2017}).

\bibitem{silvestro_ripple_2010}
\bibinfo{author}{Silvestro, S.}, \bibinfo{author}{Fenton, L.~K.},
  \bibinfo{author}{Vaz, D.~A.}, \bibinfo{author}{Bridges, N.~T.} \&
  \bibinfo{author}{Ori, G.~G.}
\newblock \bibinfo{title}{Ripple migration and dune activity on {Mars}:
  {Evidence} for dynamic wind processes}.
\newblock \emph{\bibinfo{journal}{Geophysical Research Letters}}
  \textbf{\bibinfo{volume}{37}}, \bibinfo{pages}{n/a--n/a}
  (\bibinfo{year}{2010}).

\bibitem{duran_direct_2014}
\bibinfo{author}{Duran, O.}, \bibinfo{author}{Claudin, P.} \&
  \bibinfo{author}{Andreotti, B.}
\newblock \bibinfo{title}{Direct numerical simulations of aeolian sand
  ripples}.
\newblock \emph{\bibinfo{journal}{Proceedings of the National Academy of
  Sciences}} \textbf{\bibinfo{volume}{111}}, \bibinfo{pages}{15665--15668}
  (\bibinfo{year}{2014}).

\bibitem{silvestro_dune-like_2016}
\bibinfo{author}{Silvestro, S.}, \bibinfo{author}{Vaz, D.~A.},
  \bibinfo{author}{Yizhaq, H.} \& \bibinfo{author}{Esposito, F.}
\newblock \bibinfo{title}{Dune-like dynamic of {Martian} {Aeolian} large
  ripples: Longitudinal large ripples on mars}.
\newblock \emph{\bibinfo{journal}{Geophysical Research Letters}}
  \textbf{\bibinfo{volume}{43}}, \bibinfo{pages}{8384--8389}
  (\bibinfo{year}{2016}).

\bibitem{lapotre_large_2016}
\bibinfo{author}{Lapotre, M. G.~A.} \emph{et~al.}
\newblock \bibinfo{title}{Large wind ripples on {Mars}: {A} record of
  atmospheric evolution}.
\newblock \emph{\bibinfo{journal}{Science}} \textbf{\bibinfo{volume}{353}},
  \bibinfo{pages}{55--58} (\bibinfo{year}{2016}).

\bibitem{yalin_determination_1985}
\bibinfo{author}{Yalin, M.}
\newblock \bibinfo{title}{On the determination of ripple geometry}.
\newblock \emph{\bibinfo{journal}{Journal of Hydraulic Engineering}}
  \textbf{\bibinfo{volume}{111}}, \bibinfo{pages}{1148--1155}
  (\bibinfo{year}{1985}).

\bibitem{ashley_classification_1990}
\bibinfo{author}{Ashley, G.}
\newblock \bibinfo{title}{Classification of {Large}-{Scale} {Subaqueous}
  {Bedforms}: {A} {New} {Look} at an {Old} {Problem}-{SEPM} {Bedforms} and
  {Bedding} {Structures}}.
\newblock \emph{\bibinfo{journal}{Journal of Sedimentary Research}}
  \textbf{\bibinfo{volume}{60}}, \bibinfo{pages}{160--172}
  (\bibinfo{year}{1990}).

\bibitem{andreotti_giant_2009}
\bibinfo{author}{Andreotti, B.}, \bibinfo{author}{Fourri\`ere, A.},
  \bibinfo{author}{Ould-Kaddour, F.}, \bibinfo{author}{Murray, B.} \&
  \bibinfo{author}{Claudin, P.}
\newblock \bibinfo{title}{Giant aeolian dune size determined by the average
  depth of the atmospheric boundary layer.}
\newblock \emph{\bibinfo{journal}{Nature}} \textbf{\bibinfo{volume}{457}},
  \bibinfo{pages}{1120--3} (\bibinfo{year}{2009}).

\bibitem{greeley_microdunes_1984}
\bibinfo{author}{Greeley, R.}, \bibinfo{author}{Marshall, J.~R.} \&
  \bibinfo{author}{Leach, R.~N.}
\newblock \bibinfo{title}{Microdunes and other aeolian bedforms on {Venus}:
  wind tunnel simulations}.
\newblock \emph{\bibinfo{journal}{Icarus}} \textbf{\bibinfo{volume}{60}},
  \bibinfo{pages}{152--160} (\bibinfo{year}{1984}).

\bibitem{schwammle_model_2005}
\bibinfo{author}{Schw\"ammle, V.} \& \bibinfo{author}{Herrmann, H.~J.}
\newblock \bibinfo{title}{A model of {Barchan} dunes including lateral shear
  stress.}
\newblock \emph{\bibinfo{journal}{The European physical journal. E, Soft
  matter}} \textbf{\bibinfo{volume}{16}}, \bibinfo{pages}{57--65}
  (\bibinfo{year}{2005}).

\bibitem{claudin_dissolution_2017}
\bibinfo{author}{Claudin, P.}, \bibinfo{author}{Durán, O.} \&
  \bibinfo{author}{Andreotti, B.}
\newblock \bibinfo{title}{Dissolution instability and roughening transition}.
\newblock \emph{\bibinfo{journal}{Journal of Fluid Mechanics}}
  \textbf{\bibinfo{volume}{832}} (\bibinfo{year}{2017}).

\bibitem{venditti_morphodynamics_2005}
\bibinfo{author}{Venditti, J.~G.}, \bibinfo{author}{Church, M.} \&
  \bibinfo{author}{Bennett, S.~J.}
\newblock \bibinfo{title}{Morphodynamics of small-scale superimposed sand waves
  over migrating dune bed forms}.
\newblock \emph{\bibinfo{journal}{Water Resources Research}}
  \textbf{\bibinfo{volume}{41}}, \bibinfo{pages}{n/a--n/a}
  (\bibinfo{year}{2005}).

\bibitem{venditti_interfacial_2006}
\bibinfo{author}{Venditti, J.~G.}, \bibinfo{author}{Church, M.} \&
  \bibinfo{author}{Bennett, S.~J.}
\newblock \bibinfo{title}{On interfacial instability as a cause of transverse
  subcritical bed forms}.
\newblock \emph{\bibinfo{journal}{Water Resources Research}}
  \textbf{\bibinfo{volume}{42}} (\bibinfo{year}{2006}).

\bibitem{richards_formation_1980}
\bibinfo{author}{Richards, K.~J.}
\newblock \bibinfo{title}{The formation of ripples and dunes on an erodible
  bed}.
\newblock \emph{\bibinfo{journal}{Journal of Fluid Mechanics}}
  \textbf{\bibinfo{volume}{99}}, \bibinfo{pages}{597--618}
  (\bibinfo{year}{1980}).

\bibitem{MCLEAN1990131}
\bibinfo{author}{McLean, S.~R.}
\newblock \bibinfo{title}{The stability of ripples and dunes}.
\newblock \emph{\bibinfo{journal}{Earth-Science Reviews}}
  \textbf{\bibinfo{volume}{29}}, \bibinfo{pages}{131 -- 144}
  (\bibinfo{year}{1990}).

\bibitem{andreotti_measurements_2010}
\bibinfo{author}{Andreotti, B.}, \bibinfo{author}{Claudin, P.} \&
  \bibinfo{author}{Pouliquen, O.}
\newblock \bibinfo{title}{Measurements of the aeolian sand transport saturation
  length}.
\newblock \emph{\bibinfo{journal}{Geomorphology}}
  \textbf{\bibinfo{volume}{123}}, \bibinfo{pages}{343--348}
  (\bibinfo{year}{2010}).

\bibitem{berg_new_1993}
\bibinfo{author}{Berg, J. H. V.~D.} \& \bibinfo{author}{Gelder, A.~V.}
\newblock \bibinfo{title}{A {New} {Bedform} {Stability} {Diagram}, with
  {Emphasis} on the {Transition} of {Ripples} to {Plane} {Bed} in {Flows} over
  {Fine} {Sand} and {Silt}}.
\newblock In \emph{\bibinfo{booktitle}{Alluvial {Sedimentation}}},
  \bibinfo{pages}{11--21} (\bibinfo{publisher}{Wiley-Blackwell},
  \bibinfo{year}{1993}).

\bibitem{lapotre_morphologic_2018}
\bibinfo{author}{Lapotre, M. G.~A.} \emph{et~al.}
\newblock \bibinfo{title}{Morphologic {Diversity} of {Martian} {Ripples}:
  {Implications} for {Large}-{Ripple} {Formation}}.
\newblock \emph{\bibinfo{journal}{Geophysical Research Letters}}
  (\bibinfo{year}{2018}).

\bibitem{weitz_sand_2018}
\bibinfo{author}{Weitz, C.~M.} \emph{et~al.}
\newblock \bibinfo{title}{Sand {Grain} {Sizes} and {Shapes} in {Aeolian}
  {Bedforms} at {Gale} {Crater}, {Mars}}.
\newblock \emph{\bibinfo{journal}{Geophysical Research Letters}}
  (\bibinfo{year}{2018}).

\bibitem{hugenholtz_morphology_2017}
\bibinfo{author}{Hugenholtz, C.~H.}, \bibinfo{author}{Barchyn, T.~E.} \&
  \bibinfo{author}{Boulding, A.}
\newblock \bibinfo{title}{Morphology of transverse aeolian ridges ({TARs}) on
  {Mars} from a large sample: {Further} evidence of a megaripple origin?}
\newblock \emph{\bibinfo{journal}{Icarus}} \textbf{\bibinfo{volume}{286}},
  \bibinfo{pages}{193--201} (\bibinfo{year}{2017}).

\bibitem{geissler_morphology_2017}
\bibinfo{author}{Geissler, P.~E.} \& \bibinfo{author}{Wilgus, J.~T.}
\newblock \bibinfo{title}{The morphology of transverse aeolian ridges on
  {Mars}}.
\newblock \emph{\bibinfo{journal}{Aeolian Research}}
  \textbf{\bibinfo{volume}{26}}, \bibinfo{pages}{63--71}
  (\bibinfo{year}{2017}).

\bibitem{claudin_scaling_2006}
\bibinfo{author}{Claudin, P.} \& \bibinfo{author}{Andreotti, B.}
\newblock \bibinfo{title}{A scaling law for aeolian dunes on {Mars}, {Venus},
  {Earth}, and for subaqueous ripples}.
\newblock \emph{\bibinfo{journal}{Earth and Planetary Science Letters}}
  \textbf{\bibinfo{volume}{252}}, \bibinfo{pages}{30--44}
  (\bibinfo{year}{2006}).

\bibitem{sullivan_wind-driven_2008}
\bibinfo{author}{Sullivan, R.} \emph{et~al.}
\newblock \bibinfo{title}{Wind-driven particle mobility on {Mars}: {Insights}
  from {Mars} {Exploration} {Rover} observations at “{El} {Dorado}” and
  surroundings at {Gusev} {Crater}}.
\newblock \emph{\bibinfo{journal}{Journal of Geophysical Research}}
  \textbf{\bibinfo{volume}{113}}, \bibinfo{pages}{E06S07--E06S07}
  (\bibinfo{year}{2008}).

\bibitem{bridges_formation_2015}
\bibinfo{author}{Bridges, N.~T.}, \bibinfo{author}{Spagnuolo, M.~G.},
  \bibinfo{author}{de~Silva, S.~L.}, \bibinfo{author}{Zimbelman, J.~R.} \&
  \bibinfo{author}{Neely, E.~M.}
\newblock \bibinfo{title}{Formation of gravel-mantled megaripples on {Earth}
  and {Mars}: {Insights} from the {Argentinean} {Puna} and wind tunnel
  experiments}.
\newblock \emph{\bibinfo{journal}{Aeolian Research}}
  \textbf{\bibinfo{volume}{17}}, \bibinfo{pages}{49--60}
  (\bibinfo{year}{2015}).

\bibitem{abrams_relaxation_1985}
\bibinfo{author}{Abrams, J.} \& \bibinfo{author}{Hanratty, T.~J.}
\newblock \bibinfo{title}{Relaxation effects observed for turbulent flow over a
  wavy surface}.
\newblock \emph{\bibinfo{journal}{Journal of Fluid Mechanics}}
  \textbf{\bibinfo{volume}{151}}, \bibinfo{pages}{443--455}
  (\bibinfo{year}{1985}).

\bibitem{frederick_velocity_1988}
\bibinfo{author}{Frederick, K.~A.} \& \bibinfo{author}{Hanratty, T.~J.}
\newblock \bibinfo{title}{Velocity measurements for a turbulent nonseparated
  flow over solid waves}.
\newblock \emph{\bibinfo{journal}{Experiments in Fluids}}
  \textbf{\bibinfo{volume}{6}}, \bibinfo{pages}{477--486}
  (\bibinfo{year}{1988}).

\bibitem{southard_bed_1990}
\bibinfo{author}{Southard, J.~B.} \& \bibinfo{author}{Boguchwal, L.~A.}
\newblock \bibinfo{title}{Bed configurations in steady unidirectional water
  flows. part 2: Synthesis of flume data}.
\newblock \emph{\bibinfo{journal}{Journal of Sedimentary Petrology}}
  \textbf{\bibinfo{volume}{60}}, \bibinfo{pages}{658--679}
  (\bibinfo{year}{1990}).

\bibitem{baas_predicting_2016}
\bibinfo{author}{Baas, J.~H.}, \bibinfo{author}{Best, J.~L.} \&
  \bibinfo{author}{Peakall, J.}
\newblock \bibinfo{title}{Predicting bedforms and primary current
  stratification in cohesive mixtures of mud and sand}.
\newblock \emph{\bibinfo{journal}{Journal of the Geological Society}}
  \textbf{\bibinfo{volume}{173}}, \bibinfo{pages}{12--45}
  (\bibinfo{year}{2016}).

\bibitem{andreotti_bedforms_2012}
\bibinfo{author}{Andreotti, B.}, \bibinfo{author}{Claudin, P.},
  \bibinfo{author}{Devauchelle, O.}, \bibinfo{author}{Durán, O.} \&
  \bibinfo{author}{Fourrière, A.}
\newblock \bibinfo{title}{Bedforms in a turbulent stream: ripples, chevrons and
  antidunes}.
\newblock \emph{\bibinfo{journal}{Journal of Fluid Mechanics}}
  \textbf{\bibinfo{volume}{690}}, \bibinfo{pages}{94--128}
  (\bibinfo{year}{2012}).

\bibitem{lammel_aeolian_2018}
\bibinfo{author}{Lämmel, M.} \emph{et~al.}
\newblock \bibinfo{title}{Aeolian sand sorting and megaripple formation}.
\newblock \emph{\bibinfo{journal}{Nature Physics}}
  \textbf{\bibinfo{volume}{14}}, \bibinfo{pages}{759--765}
  (\bibinfo{year}{2018}).

\bibitem{duran_numerical_2012}
\bibinfo{author}{Dur\'an, O.}, \bibinfo{author}{Andreotti, B.} \&
  \bibinfo{author}{Claudin, P.}
\newblock \bibinfo{title}{Numerical simulation of turbulent sediment transport,
  from bed load to saltation}.
\newblock \emph{\bibinfo{journal}{Physics of Fluids}}
  \textbf{\bibinfo{volume}{24}}, \bibinfo{pages}{103306--103306}
  (\bibinfo{year}{2012}).

\bibitem{ungar_steady_1987}
\bibinfo{author}{Ungar, J.~E.} \& \bibinfo{author}{Haff, P.~K.}
\newblock \bibinfo{title}{Steady state saltation in air}.
\newblock \emph{\bibinfo{journal}{Sedimentology}}
  \textbf{\bibinfo{volume}{34}}, \bibinfo{pages}{289--299}
  (\bibinfo{year}{1987}).

\bibitem{marshall_experimental_1992}
\bibinfo{author}{Marshall, J.~R.} \& \bibinfo{author}{Greeley, R.}
\newblock \bibinfo{title}{An experimental study of aeolian structures on
  {Venus}}.
\newblock \emph{\bibinfo{journal}{Journal of Geophysical Research: Planets}}
  \textbf{\bibinfo{volume}{97}}, \bibinfo{pages}{1007--1016}
  (\bibinfo{year}{1992}).

\bibitem{pahtz_flux_2013}
\bibinfo{author}{P\"ahtz, T.}, \bibinfo{author}{Kok, J.~F.},
  \bibinfo{author}{Parteli, E. J.~R.} \& \bibinfo{author}{Herrmann, H.~J.}
\newblock \bibinfo{title}{Flux {Saturation} {Length} of {Sediment}
  {Transport}}.
\newblock \emph{\bibinfo{journal}{Physical Review Letters}}
  \textbf{\bibinfo{volume}{111}}, \bibinfo{pages}{218002}
  (\bibinfo{year}{2013}).

\bibitem{pahtz_fluid_2017}
\bibinfo{author}{P\"ahtz, T.} \& \bibinfo{author}{Dur\'an, O.}
\newblock \bibinfo{title}{Fluid forces or impacts: {What} governs the
  entrainment of soil particles in sediment transport mediated by a {Newtonian}
  fluid?}
\newblock \emph{\bibinfo{journal}{Physical Review Fluids}}
  \textbf{\bibinfo{volume}{2}}, \bibinfo{pages}{074303} (\bibinfo{year}{2017}).

\bibitem{sullivan_aeolian_2017}
\bibinfo{author}{Sullivan, R.} \& \bibinfo{author}{Kok, J.~F.}
\newblock \bibinfo{title}{Aeolian saltation on {Mars} at low wind speeds}.
\newblock \emph{\bibinfo{journal}{Journal of Geophysical Research: Planets}}
  \textbf{\bibinfo{volume}{122}} (\bibinfo{year}{2017}).

\bibitem{pont_two_2014}
\bibinfo{author}{Pont, S. C.~d.}, \bibinfo{author}{Narteau, C.} \&
  \bibinfo{author}{Gao, X.}
\newblock \bibinfo{title}{Two modes for dune orientation}.
\newblock \emph{\bibinfo{journal}{Geology}} \textbf{\bibinfo{volume}{42}},
  \bibinfo{pages}{743--746} (\bibinfo{year}{2014}).

\bibitem{guy_summary_1966}
\bibinfo{author}{Guy, H.~P.}, \bibinfo{author}{Simons, D.~B.} \&
  \bibinfo{author}{Richardson, E.~V.}
\newblock \emph{\bibinfo{title}{Summary of {Alluvial} {Channel} {Data} from
  {Flume} {Experiments}, 1956-61}} (\bibinfo{publisher}{U.S. Government
  Printing Office}, \bibinfo{year}{1966}).

\bibitem{martin_origin_2013}
\bibinfo{author}{Martin, R.~L.} \& \bibinfo{author}{Jerolmack, D.~J.}
\newblock \bibinfo{title}{Origin of hysteresis in bed form response to unsteady
  flows}.
\newblock \emph{\bibinfo{journal}{Water Resources Research}}
  \textbf{\bibinfo{volume}{49}}, \bibinfo{pages}{1314--1333}
  (\bibinfo{year}{2013}).

\bibitem{baas_flume_1994}
\bibinfo{author}{Baas, J.~H.}
\newblock \bibinfo{title}{A flume study on the development and equilibrium
  morphology of current ripples in very fine sand}.
\newblock \emph{\bibinfo{journal}{Sedimentology}}
  \textbf{\bibinfo{volume}{41}}, \bibinfo{pages}{185--209}
  (\bibinfo{year}{1994}).

\bibitem{baas_empirical_1999}
\bibinfo{author}{Baas, J.~H.}
\newblock \bibinfo{title}{An empirical model for the development and
  equilibrium morphology of current ripples in fine sand}.
\newblock \emph{\bibinfo{journal}{Sedimentology}}
  \textbf{\bibinfo{volume}{46}}, \bibinfo{pages}{123--138}
  (\bibinfo{year}{1999}).

\bibitem{langlois_initiation_2007}
\bibinfo{author}{Langlois, V.} \& \bibinfo{author}{Valance, A.}
\newblock \bibinfo{title}{Initiation and evolution of current ripples on a flat
  sand bed under turbulent water flow}.
\newblock \emph{\bibinfo{journal}{The European Physical Journal E}}
  \textbf{\bibinfo{volume}{22}}, \bibinfo{pages}{201--208}
  (\bibinfo{year}{2007}).

\bibitem{coleman_initiation_1996}
\bibinfo{author}{Coleman, S.~E.} \& \bibinfo{author}{Melville, B.~W.}
\newblock \bibinfo{title}{Initiation of bed forms on a flat sand bed}.
\newblock \emph{\bibinfo{journal}{Journal of hydraulic engineering}}
  \textbf{\bibinfo{volume}{122}}, \bibinfo{pages}{301--310}
  (\bibinfo{year}{1996}).

\bibitem{kuru_formation_1995}
\bibinfo{author}{Kuru, W.~C.}, \bibinfo{author}{Leighton, D.~T.} \&
  \bibinfo{author}{McCready, M.~J.}
\newblock \bibinfo{title}{Formation of waves on a horizontal erodible bed of
  particles}.
\newblock \emph{\bibinfo{journal}{International Journal of Multiphase Flow}}
  \textbf{\bibinfo{volume}{21}}, \bibinfo{pages}{1123--1140}
  (\bibinfo{year}{1995}).

\bibitem{jain_growth_1971}
\bibinfo{author}{Jain, S.~C.} \& \bibinfo{author}{Kennedy, J.~F.}
\newblock \bibinfo{title}{The growth of sand waves}.
\newblock In \emph{\bibinfo{booktitle}{Proc. {Int}. {Symp}. on {Stochastic}
  {Hydr}.}}, \bibinfo{pages}{449--471} (\bibinfo{publisher}{Pittsburgh
  University Press}, \bibinfo{address}{Pittsburgh}, \bibinfo{year}{1971}).

\bibitem{jain_spectral_1974}
\bibinfo{author}{Jain, S.~C.} \& \bibinfo{author}{Kennedy, J.~F.}
\newblock \bibinfo{title}{The spectral evolution of sedimentary bed forms}.
\newblock \emph{\bibinfo{journal}{Journal of Fluid Mechanics}}
  \textbf{\bibinfo{volume}{63}}, \bibinfo{pages}{301--314}
  (\bibinfo{year}{1974}).

\bibitem{nakagawa_hiroji_spectral_1984}
\bibinfo{author}{{Nakagawa Hiroji}} \& \bibinfo{author}{{Tsujimoto Tetsuro}}.
\newblock \bibinfo{title}{Spectral {Analysis} of {Sand} {Bed} {Instability}}.
\newblock \emph{\bibinfo{journal}{Journal of Hydraulic Engineering}}
  \textbf{\bibinfo{volume}{110}}, \bibinfo{pages}{467--483}
  (\bibinfo{year}{1984}).

\bibitem{bridge_flow_1988}
\bibinfo{author}{Bridge, J.~S.} \& \bibinfo{author}{Best, J.~L.}
\newblock \bibinfo{title}{Flow, sediment transport and bedform dynamics over
  the transition from dunes to upper-stage plane beds: implications for the
  formation of planar laminae}.
\newblock \emph{\bibinfo{journal}{Sedimentology}}
  \textbf{\bibinfo{volume}{35}}, \bibinfo{pages}{753--763}
  (\bibinfo{year}{1988}).

\bibitem{greeley_windblown_1984}
\bibinfo{author}{Greeley, R.} \emph{et~al.}
\newblock \bibinfo{title}{Windblown sand on {Venus}: {Preliminary} results of
  laboratory simulations}.
\newblock \emph{\bibinfo{journal}{Icarus}} \textbf{\bibinfo{volume}{57}},
  \bibinfo{pages}{112--124} (\bibinfo{year}{1984}).

\bibitem{kok_difference_2010}
\bibinfo{author}{Kok, J.~F.}
\newblock \bibinfo{title}{Difference in the {Wind} {Speeds} {Required} for
  {Initiation} versus {Continuation} of {Sand} {Transport} on {Mars}:
  {Implications} for {Dunes} and {Dust} {Storms}}.
\newblock \emph{\bibinfo{journal}{Physical Review Letters}}
  \textbf{\bibinfo{volume}{104}} (\bibinfo{year}{2010}).

\bibitem{newman_winds_2017}
\bibinfo{author}{Newman, C.~E.} \emph{et~al.}
\newblock \bibinfo{title}{Winds measured by the {Rover} {Environmental}
  {Monitoring} {Station} ({REMS}) during the {Mars} {Science} {Laboratory}
  ({MSL}) rover's {Bagnold} {Dunes} {Campaign} and comparison with numerical
  modeling using {MarsWRF}}.
\newblock \emph{\bibinfo{journal}{Icarus}} \textbf{\bibinfo{volume}{291}},
  \bibinfo{pages}{203--231} (\bibinfo{year}{2017}).

\bibitem{andreotti_aeolian_2006}
\bibinfo{author}{Andreotti, B.}, \bibinfo{author}{Claudin, P.} \&
  \bibinfo{author}{Pouliquen, O.}
\newblock \bibinfo{title}{Aeolian {Sand} {Ripples}: {Experimental} {Study} of
  {Fully} {Developed} {States}}.
\newblock \emph{\bibinfo{journal}{Physical Review Letters}}
  \textbf{\bibinfo{volume}{96}}, \bibinfo{pages}{028001--028001}
  (\bibinfo{year}{2006}).

\end{thebibliography}

\begin{addendum}
	\item[Corresponding author] Correspondence and requests for materials should be addressed to O.D.V. (oduranvinent@tamu.edu).
	\item[Acknowledgements] O.D.V. and C.W. were funded through the DFG Research Center/Cluster of Excellence ``The Ocean in the Earth System''.
 	\item[Author contributions] O.D.V. and C.W. contributed to the conception of the work and the analysis of bedform data. O.D.V. performed the simulations. O.D.V, B.A. and P.C. contributed to the validation of the model, interpretation of the results and writing of the manuscript.
 	\item[Competing Interests] The authors declare no competing financial interests.
	\item[Data availability] The authors declare that the data supporting the findings of this study are within the corresponding references and available from the authors upon request. Some of the data is also within the supplementary information file.
	\item[Code availability] The code integrating the model equations used for this study can be made available upon request from the authors.
\end{addendum}
\newpage

%%%% FIGURE CAPTIONS %%%%%

%
\begin{figure}
 \begin{center}
  \includegraphics{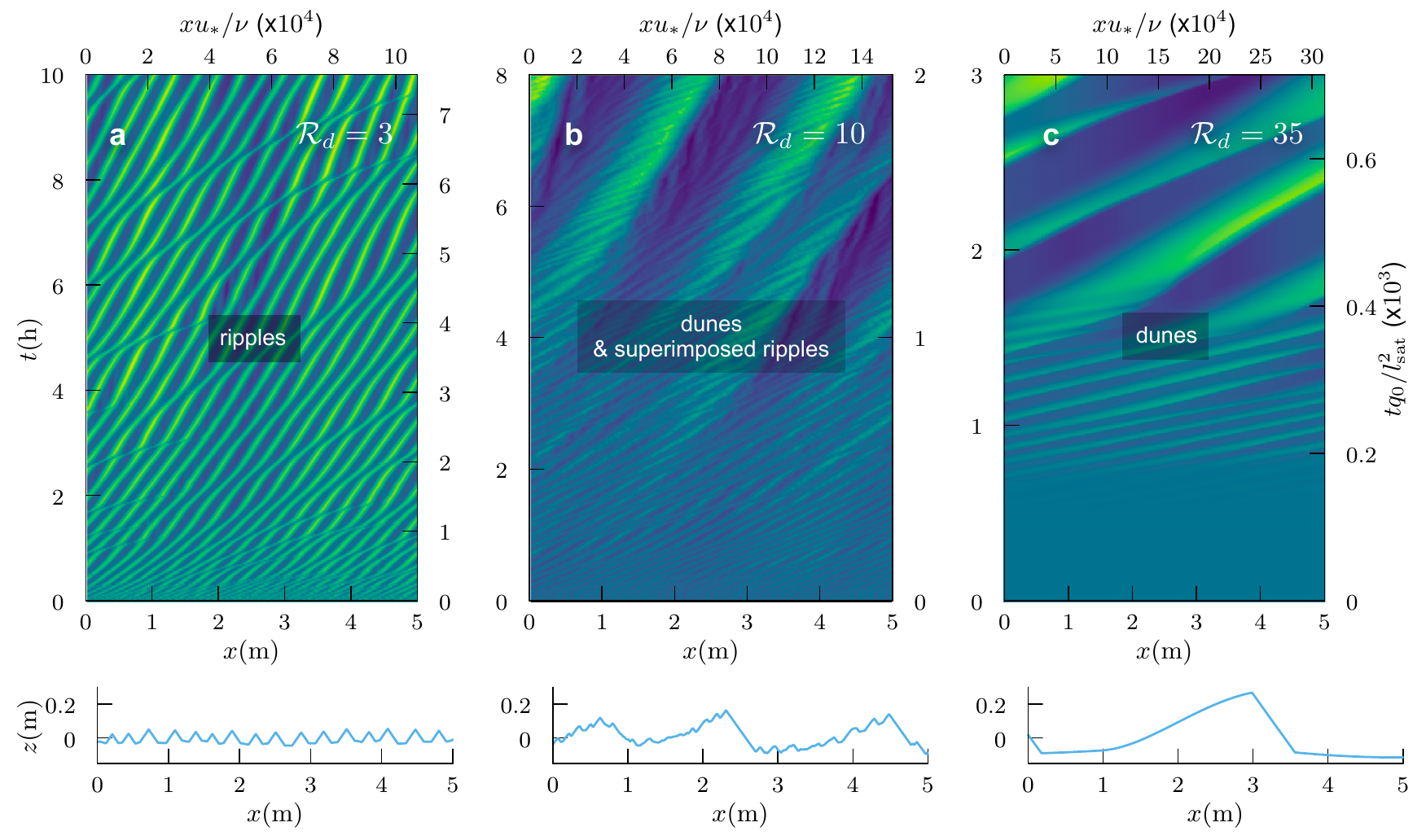}
 \end{center}
 \caption{ \small{ {\bf Temporal evolution of simulated subaqueous bedforms.} Simulations for three different values of the grain-based Reynolds number $\mathcal{R}_d \approx u_* d/\nu$ corresponding, following the current classification for water bedforms, to ripples ({\bf a}), dunes with superimposed ripples ({\bf b}), and dunes ({\bf c}) (see also Supplementary Movies 1-3). ({\bf d-e}) Mature profile for each condition. For the simulations we used $u_*/u_t = 2.2$ ({\bf a}), $2.5$ ({\bf b}) and $4.2$ ({\bf c}). These values were selected to give a typical value for the saturation length of bedload transport ($l_{\rm sat} \approx 42d$, see Supp.Fig.~5) when using the formulation of $l_{\rm sat}$ described in the Methods. In panels ({\bf a-c}), both dimensional (left and bottom) and dimensionless (right and top) axes are shown. In the time rescaling factor, $q_0 = Q [(u_*/u_t)^2-1] (u_*/u_t)$ is the average sand flux, where $Q$ is the flux scale defined in the Methods.} }
 %\label{Fig-SpatioTemporal}
\end{figure}
\begin{figure}
 \begin{center}
  \includegraphics{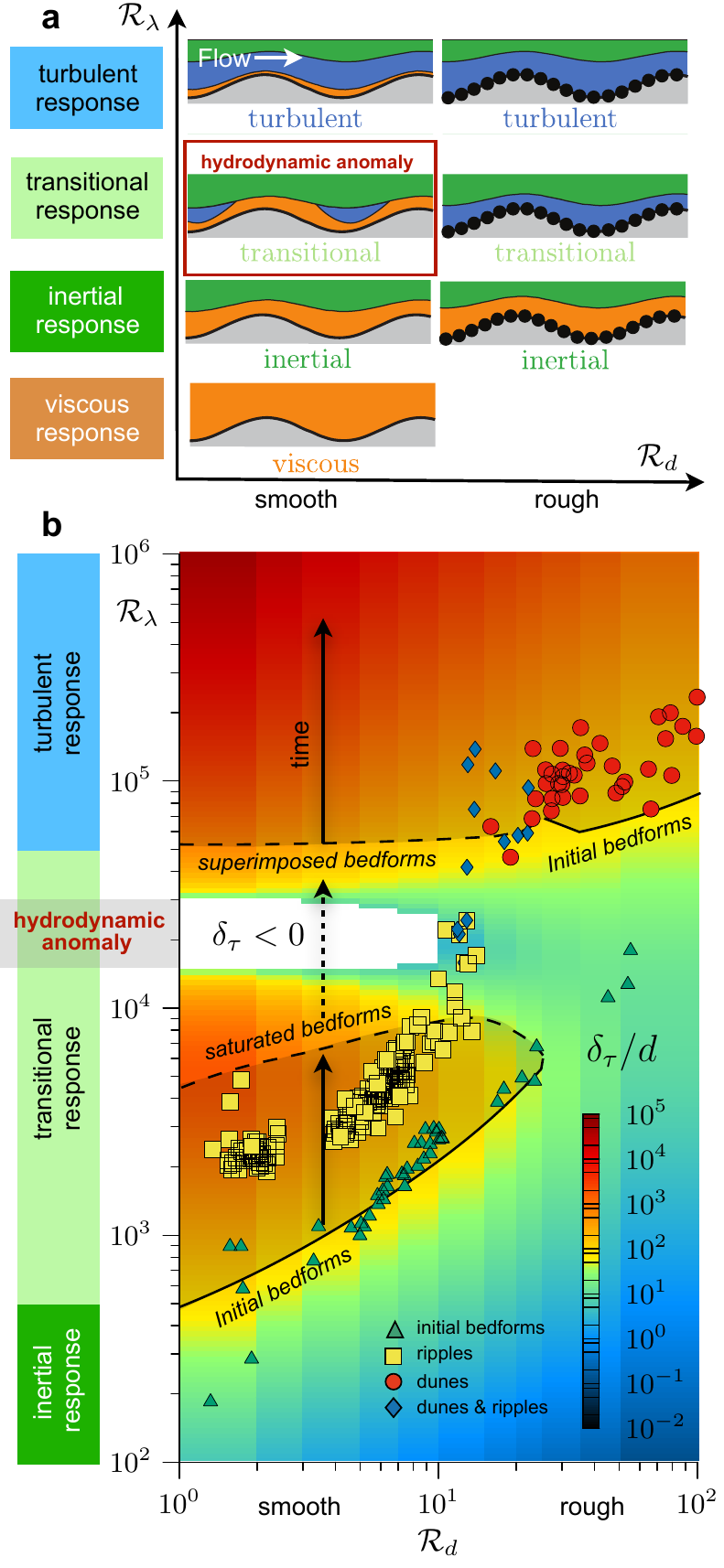}
 \end{center}
 \caption{ \small{ {\bf Hydrodynamic mechanisms and predictions for subaqueous bedforms.} 
({\bf a}) Schematics of the layered hydrodynamic response to the bed topography as function of $\mathcal{R}_\lambda = \lambda u_*/\nu$ and $\mathcal{R}_d = u_* d/\nu$. Colours represent the dominant hydrodynamic mechanism balancing the pressure gradient modulation: viscous diffusion (orange), inertia (green) or turbulent fluctuations (blue). ({\bf b}) Rescaled shear stress spatial shift $\delta_\tau/d$ (in color). Hydrodynamic anomaly is shown in white. Shaded areas are predicted wavelengths for a typical bedload saturation length $l_{\rm sat}/d = 42$ (Supp.Fig.~5). Symbols are experimental data for monodisperse sand in water or water-glycerin mixtures with $l_{\rm sat}/d$ in the range $30-60$ (see Methods and Supp.Fig.~13 for data sources). Data is reported as ripples, dunes or dunes with superimposed ripples. Solid lines show the most unstable mode (initial bedforms). Dashed-lines represent the modes emerging from the non-linear dynamics: saturated bedforms and superimposed bedforms. Solid arrows represent temporal pattern coarsening while the dashed arrow represents a jump in the dominant wavelength. }}
% \label{Fig-water}
\end{figure}
\begin{figure}
 \begin{center}
  \includegraphics{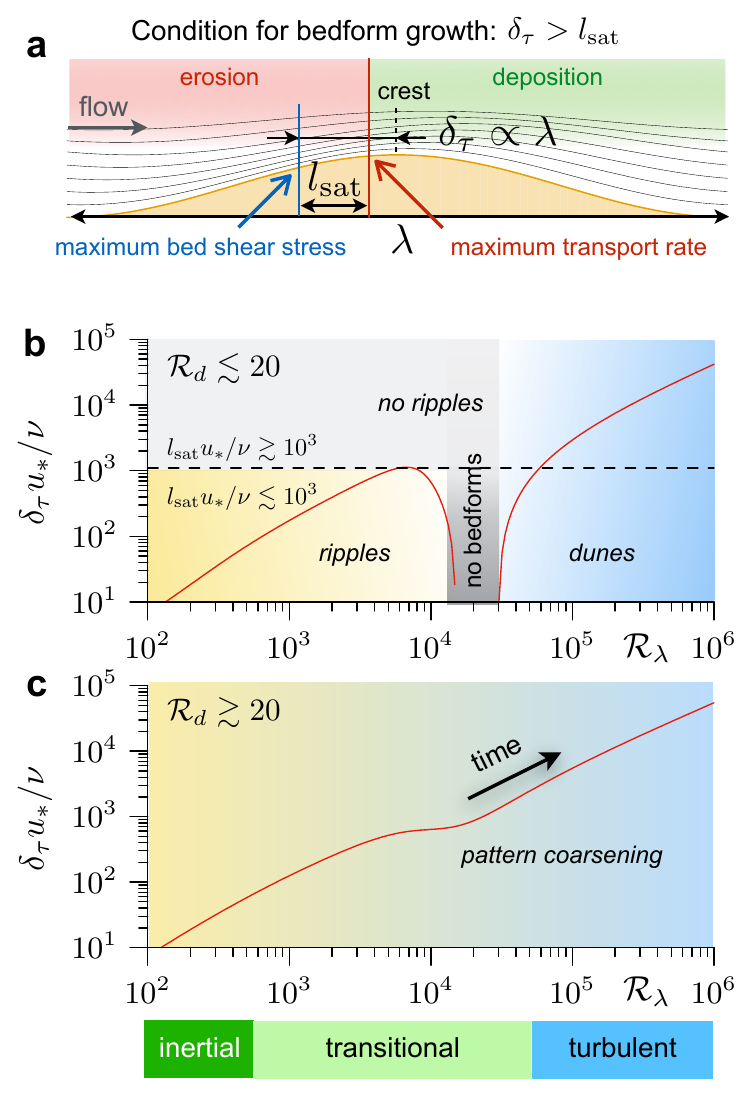}
 \end{center}
 \caption{ \small{ {\bf Effects of the saturation length on bedform formation.} ({\bf a}) 
Necessary condition for bedform growth with an sketch defining the shear stress spatial shift $\delta_\tau$ and the transport saturation length $l_{\rm sat}$ (see Methods for a precise definition). ({\bf b-c}) Rescaled shear stress shift as function of $\mathcal{R}_d$ for hydrodynamically smooth ($\mathcal{R}_d = 1$, {\bf b}) and rough ($\mathcal{R}_d = 50$, {\bf c}) flows. ({\bf b}) Smooth or transitional flows: both ripples and dunes are possible for $l_{\rm sat} u_*/\nu \lesssim 10^3$; otherwise ripples disappear. ({\bf c}) Rough flows: there is no hydrodynamic anomaly and bedforms increase their size by pattern coarsening. Therefore, there are no stable ripples. Color bands represent the different hydrodynamic responses induced by the bedforms (Fig.~3a). The local maximum of $\delta_\tau u_*/\nu$ in ({\bf b}) changes by less than a factor of 2 in the range $\mathcal{R}_d \lesssim 20$ (data not shown).} }
% \label{Fig-mech}
\end{figure}
\begin{figure}
 \begin{center}
  \includegraphics[width=1\textwidth]{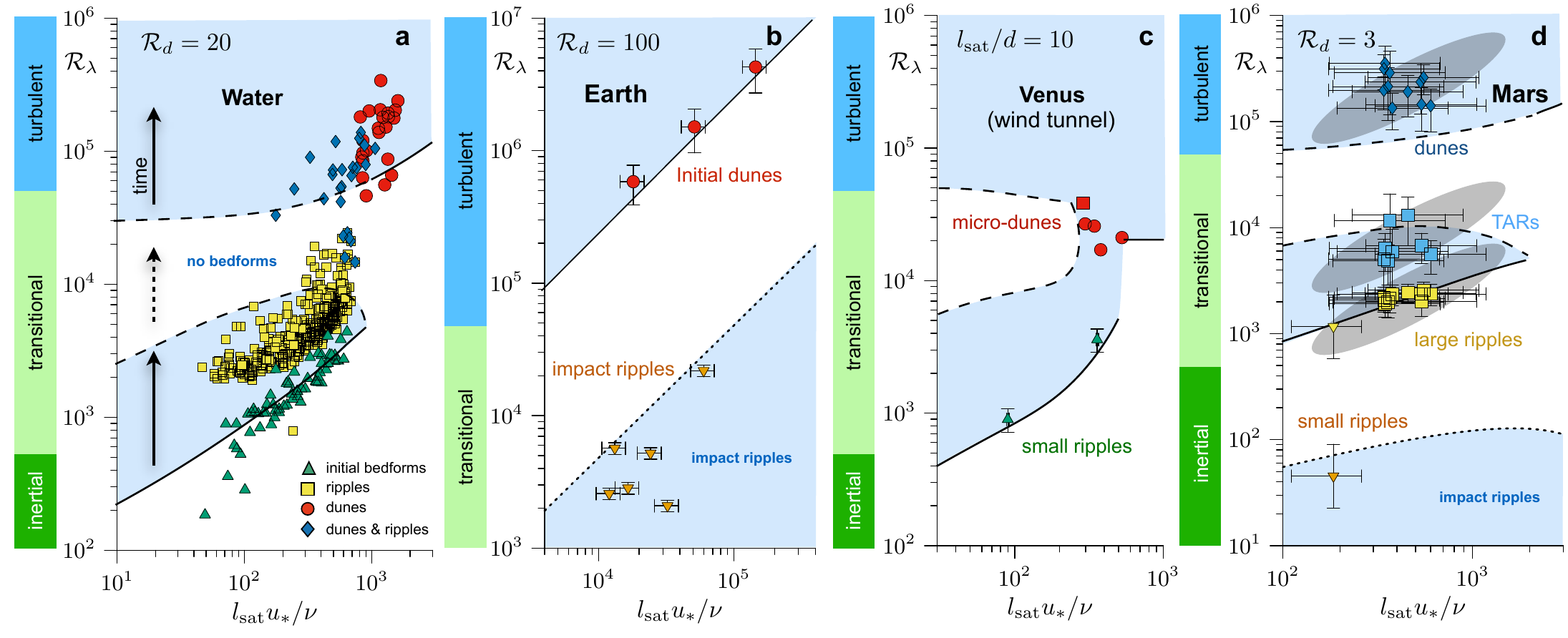}
 \end{center}
 \caption{ \small{ {\bf Predictions for different planetary conditions and transport modes.} Predicted rescaled wavelengths $\mathcal{R}_\lambda = \lambda u_*/\nu$ (blue areas) as function of the rescaled saturation length $l_{\rm sat} u_*/\nu$, for a characteristic $\mathcal{R}_d$ corrected to include transport effects (see Methods and Supp.Fig.~18). Solid lines are the most unstable modes. Dashed lines are the modes emerging from the non-linear evolution (see Fig.~2). Dotted lines are semi-empirical estimations of the typical maximum size of impact ripples for monodisperse sand (Methods). Solid arrows represent temporal pattern coarsening. The dashed arrow represents a jump in the dominant wavelength. Symbols are bedform data for monodisperse sand (except for Martian TARs---transversal aeolian ridges---see text) described as reported. ({\bf a}) Water and water-glycerin mixtures: experimental data in the range $\mathcal{R}_d = 1-20$ (see Methods and Supp.Fig.~16 for data sources and Supp.Fig.~5 for the uncertainty range). ({\bf b}) Earth: initial dunes\cite{andreotti_measurements_2010} (Supp.Table~2) and impact ripples\cite{andreotti_aeolian_2006}. ({\bf c}) Venus: wind tunnel bedform data ($\circ$\cite{greeley_microdunes_1984}, $\triangle$\cite{greeley_microdunes_1984} and $\square$\cite{marshall_experimental_1992}, Supp.Table~3). Model predictions for a calibrated value $l_{\rm sat}/d = 10$. ({\bf d}) Mars: orbital bedform data\cite{lapotre_large_2016} ($\square$ and $\diamond$, Supp.Table~4) and in-situ data from the Namib dune, Gale Crater\cite{lapotre_large_2016} ($\triangledown$, Supp.Table~5). Horizontal errorbars reflect the uncertainty on $l_{\rm sat}$ and shadow areas represent the uncertainty in $u_*$ (Methods and Supp.Table~6). }}
% \label{Fig-mars}
\end{figure}
\begin{figure}
 \begin{center}
  \includegraphics{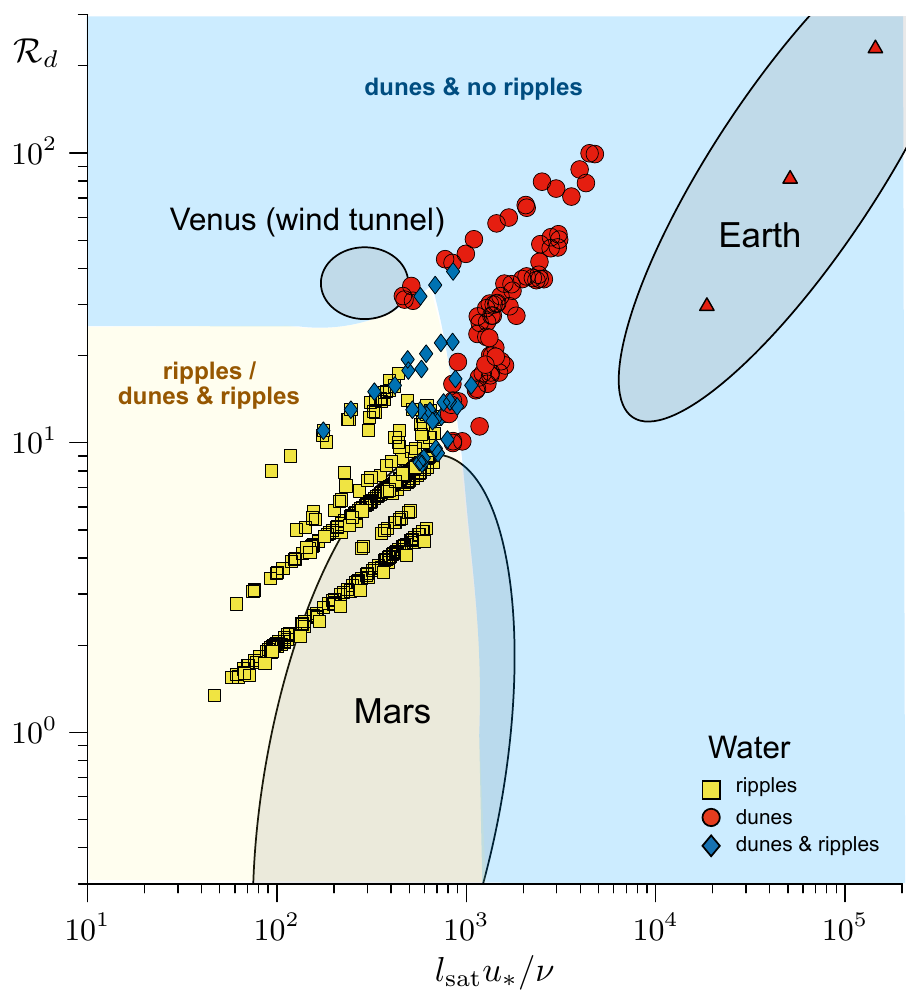}
 \end{center}
 \caption{ \small{ {\bf Phase diagram of ripples and dunes.} Presence of ripples and/or dunes as function of the rescaled saturation length $l_{\rm sat} u_*/\nu$ and grain-based Reynolds number $\mathcal{R}_d$ corrected to include transport effects (Methods). The two distinct regimes are defined by the presence or absence of stable ripples. In yellow, conditions leading to the presence of stable (saturated) ripples, with or without superimposed dunes. In blue, conditions leading to the absence of stable ripples and thus to only dunes. Symbols corresponds to water and water/glycerin mixture data ($\square$, $\circ$ and $\diamond$), and aeolian (Earth) data ($\triangle$), both for monodisperse sand. Shadow areas correspond to ranges for different planetary conditions and follow data from Fig.~4 and Supp.Tables~2,3 and 6. The boundary between the two regimes is approximately defined by the conditions: $l_{\rm sat} u_*/d \approx 10^3$ and $\mathcal{R}_d \approx 25$.} } 
% \label{Fig-regimes}
\end{figure}
%

%%
%% TABLES
%%

%

\setcounter{equation}{0}
\setcounter{figure}{0}
\renewcommand{\theequation}{S\arabic{equation}}
\renewcommand{\thefigure}{(Extended Data) \arabic{figure}}

%\newpage
%\section*{Supplementary Online Text}

\end{document}